# The Weighted Maximum-Mean Subtree and Other Bicriterion Subtree Problems

Josiah Carlson and David Eppstein

Computer Science Department
University of California, Irvine CA 92697, USA,
{jcarlson,eppstein}@uci.edu

**Abstract.** We consider problems in which we are given a rooted tree as input, and must find a subtree with the same root, optimizing some objective function of the nodes in the subtree. When this function is the sum of constant node weights, the problem is trivially solved in linear time. When the objective is the sum of weights that are linear functions of a parameter, we show how to list all optima for all possible parameter values in $O(n \log n)$ time; this parametric optimization problem can be used to solve many bicriterion optimizations problems, in which each node has two values $x_i$ and $y_i$ associated with it, and the objective function is a bivariate function $f(\sum x_i, \sum y_i)$ of the sums of these two values. A special case, when $f$ is the ratio of the two sums, is the Weighted Maximum-Mean Subtree Problem, or equivalently the Fractional Prize-Collecting Steiner Tree Problem on Trees; for this special case, we provide a linear time algorithm for this problem when all weights are positive, improving a previous $O(n \log n)$ solution, and prove that the problem is NP-complete when negative weights are allowed.

## 1 Introduction

Suppose we are given a rooted tree, in which each node $i$ has two quantities $x_i$ and $y_i$ associated with it, and a bivariate objective function $f$. For any subtree $S$, let $X_S = \sum_{i \in S} x_i$ and $Y_S = \sum_{i \in S} y_i$. We wish to find the subtree $S$, having the same root as the input tree, that maximizes $f(X_S, Y_S)$. (We note that finding the tree minimizing or maximizing $\sum x_i$ is a simple variant of the Open Pit Mining problem on DAGs [2], and can be easily solved in $O(n)$ time [7].)

For instance, if $f(X, Y) = X/Y$, we can interpret $x_i$ as the *profit* of a node, and $y_i$ as the *cost* of a node; the optimal subtree is the one that maximizes the return on investment. This problem, which we call the Weighted Maximum-Mean Subtree Problem (WMMSTP), can also be viewed as a special case of the Fractional Prize-Collecting Steiner Tree Problem (FPCSTP); in the FPCSTP, one is given a graph, with costs on the edges and profits on the vertices, a starting vertex $v_0$, and a starting cost $c_0$, and must find a tree rooted at $v_0$ that maximizes the total profit divided by the total cost. It is easy to see that the WMMSTP is equivalent to a special case of the FPCSTP in which the input is a rooted tree, $v_0$ is the root of the tree, $c_0$ is the cost of the root node, and each additional tree node has a cost in the WMMSTP that is equal to the cost in the FPCSTP of the edge connecting the node to its parent. For this special case of FPCSTP on Trees (or equivalently WMMSTP), when all costs are positive, an $O(n \log n)$ time algorithm is known [7].

The Weighted Maximum-Mean Subtree Problem (and equivalently the FPCSTT problem) can be applied to the problem of maximizing return on investment for adding services to a preexisting tree structured utility networks. An example of this lies within the development of DSL services over preexisting telephone networks, as has been in construction in recent years. The costs associated with adding such a service involve additional equipment (repeaters, hubs, switches, filters, and line upgrades) placed along the standard wired telephone services, and profits are simply the expected profits gained from providing such a service to homes and businesses connected to the upgraded network. In running the algorithm we present, a telephone company that desires to upgrade their lines would discover where they should offer such services so as to maximize the percent return on their service upgrade investment. More generally this problem applies to any situation in which return on investment is to be optimized among a tree-structured or hierarchical set of business opportunities.

More generally let $g(X, Y) = X + \lambda Y$, and consider the sequence of subtrees that maximize $g(X_S, Y_S)$ as the parameter $\lambda$ varies from $-\infty$ to $+\infty$. We call the problem of computing this sequence the Parametric Linear Maximum Subtree Problem. This sequence of parametric optimal subtrees can be viewed as forming the upper convex hull of the planar point set formed by taking one point $(X_S, Y_S)$ for each possible subtree $S$ of the input tree, and the lower convex hull can be formed similarly as the sequence of trees minimizing $g(X_S, Y_S)$ as $\lambda$ varies.

As has been previously seen for related bicriterion spanning tree problems [1, 5, 6], many bicriterion optimal subtree problems can be solved by this parametric approach. If $f(X, Y)$ is any convex or quasiconvex function, its maximum over the set of points $(X_S, Y_S)$ is achieved at a vertex of the convex hull, so by computing and testing all parametric optima we can be certain to find the subtree that maximizes a convex or quasiconvex $f(X, Y)$ or equivalently that minimizes a concave or quasiconcave $f(X, Y)$. The following problems can all be solved with this parametric formulation:

- The Weighted Maximum-Mean Subtree Problem can be viewed as a special case of this formulation in which $f(X, Y) = X/Y$; this function is convex on the upper halfplane (positive costs) but not on the whole plane (where costs may be negative). We supply a more efficient algorithm for this problem when all weights are positive, but the parametric approach succeeds more generally when weights may be negative but all subtrees containing the tree root have positive total weight.
- Suppose node $i$ may fail with probability $p_i$ and has a cost $x_i$, and let $y_i = -\ln(1 - p_i)$. Then the reliability probability of the overall subtree is $e^Y$, and the subtree that minimizes the ratio of cost to reliability can be found by minimizing $f(X, Y) = Xe^Y$.
- If nodes have weights that are unknown random variables with known mean and variance, the stochastic programming problem of finding a subtree with high probability of having low weight can be expressed as minimizing $f(X, Y) = X + \sqrt{Y}$, and the problem of finding a subtree minimizing the variance in weight can be expressed as minimizing $f(X, Y) = X - Y^2$.

As we will show, this parametric approach leads to efficient algorithms for these and many other bicriterion optimal subtree problems.

Our algorithms assume that the values $x_i$ and $y_i$ associated with tree nodes are real-valued, and that any single arithmetic operation or comparison on two real values can be performed in constant time.

## 2 New Results

We provide the following results:

- A linear time algorithm for solving the Weighted Maximum-Mean Subtree Problem with positive weights, improving a previous $O(n \log n)$ time solution [7].
- A proof that the Weighted Maximum-Mean Subtree Problem is NP-complete when the weights are allowed to be negative.

- An optimal $O(n \log n)$ time algorithm for listing all solutions, in order by parameter value, to the Parametric Linear Maximum Subtree Problem.
- An $O(n \log n)$ time algorithm for solving any bicriterion optimal subtree problem that maximizes a convex function $f(X, Y)$ or minimizes a concave function $f(X, Y)$ of the sums $X$ and $Y$ of the two node values.

## 3 The Weighted Maximum-Mean Subtree Problem

We are given a rooted tree of nodes such that each node $i$ has a real valued *profit* $x_i$, and are to produce a subtree that maximizes the average profit of the remaining nodes. By *pruning a node* we mean removing it and all of its descendants from the input; our task can then be phrased as finding an optimal set of nodes to prune.

A generalization of this problem gives each node a positive real valued *cost* $y_i$; the original problem can be viewed as assigning each node a unit cost. The overall average of a tree is the sum of the profits divided by the sum of the costs, including only unpruned nodes. In this section we show the problem to be NP-complete when the costs can be negative, and present an algorithm that solves the generalization with both profits and positive costs per node in time $O(n)$.

We assume that our input consists of a rooted tree $T$, whose nodes have positive or negative real valued profits and positive real valued costs on each node. The output should be a pruned subtree $P(T)$ and an average profit

$$OPTAVG = AVG(P(T)) = \sum_{x \in U} profit(x) / \sum_{x \in U} cost(x),$$

where $U$ denotes the set of unpruned nodes in $P(T)$, and where this average profit is at least as large as that of any other subtree of $T$ having the same root.

### 3.1 NP-Completeness with Negative Costs

**Theorem 1.** *The Weighted Maximum-Mean Subtree Problem with negative cost nodes is NP-complete.*

*Proof.* Given an instance of Integer Subset-Sum with a set $S$ of values, and a desired total $U$, we create a rooted tree with a root node and $|S|$ leaf nodes hanging from the root. Set the root's profit to one, the root's cost to $U$, and all leaf profits to zero. Assign each leaf's cost as the negation of one of the values from $S$. The maximum mean for such a tree is $\infty$, if and only if the Subset Sum instance has a subset of values summing to $U$, where in this case the optimal subtree includes only those leaf nodes whose costs sum to $U$. ∎

By slightly adjusting the root cost in this reduction we can avoid the issue of division by zero while still preserving the computational complexity of the problem. Due to this result, in the rest of the section we restrict costs to being strictly positive. However, our algorithm correctly finds the maximum mean subtree even when profits are allowed to be negative.

```
def HasAverageAtLeast(tree, cutoff):
    tree.subprofit = tree.profit
    tree.subcost = tree.cost

    for child in tree.children:
        if HasAverageAtLeast(child, cutoff):
            tree.subprofit += child.subprofit
            tree.subcost += child.subcost

    unpruned = tree.subprofit/tree.subcost >= cutoff
    tree.pruned = not unpruned
    return unpruned
```

Listing 1: Testing whether $OPTAVG$ is at least a given cutoff.

## 3.2 Two Decision Algorithms

We first define the algorithm provided in Listing 1 which tells us whether or not some tree has a pruning with average greater than or equal to some provided cutoff. Essentially, the algorithm traverses the tree bottom-up, pruning any node when the average of it and its unpruned descendants falls below the cutoff.

**Lemma 1.** *Suppose that there exists a tree $T$ with average value at least cutoff. Then the tree $U$ that $HasAverageAtLeast$ forms by pruning the input tree also has average value at least cutoff.*

*Proof.* $T$ and $U$ may differ, by the inclusion of some subtrees and the exclusion of others. For each subtree $s$ that is included in $U$ and excluded from $T$, $s$ must have average value at least cutoff (otherwise, $HasAverageAtLeast$ would have pruned it) so combining its value from that of $T$ can not bring the average below cutoff. For each subtree $s$ that is excluded from $U$ and included in $T$, $s$ must have average value below cutoff (by an inductive application of the lemma to the subtree rooted at the root of $s$) so removing its value from that of $T$ can only increase the overall value. ∎

**Corollary 1.** *$HasAverageAtLeast$ returns True if and only if there exists a tree with average value at least cutoff.*

A very similar subroutine, which we call $HasAverageGreaterThan$, replaces the greater-or-equal test in the assignment to *unpruned* by a strict inequality, and a suitably modified version of Lemma 1 and its proof would also prove its correctness.

The *tree.subprofit*, *tree.subcost*, and *tree.pruned* variables provided above are implementation details. In our implementation, *tree.subprofit* and *tree.subcost* allow for the caching of the sum of unpruned node profits and costs in a subtree rooted at *tree*, and *tree.pruned* allows for the generation of a pruned version of a tree as a side-effect of determining whether or not a tree has a pruning of average at least or greater than the specified cutoff.

## 3.3 Previous Maximum-Mean Subtree Algorithms

Klau *et al.* [7] provide three algorithms which solve the Maximum-Mean Subtree problem. The first is a simple binary search, based on the linear time decision algorithms presented above, which runs in $O(nk)$ time, where $k$ denotes the desired precision of the answer in bits, and $n$ is the number of nodes in the input tree. A second algorithm, which they present as a form of Newton's Method, runs in worst-case $O(n^2)$ time on a tree with $n$ nodes. Their third algorithm, which they present as their main result, is based on Megiddo's Parametric Search and runs in $O(n \log n)$ time.

We briefly describe this third algorithm, as our linear time algorithm is closely related to it. It performs a sequence of iterations, each of which performs a binary search among the profit/cost ratios of the remaining tree nodes to determine their relation to the optimal solution value. Once all of these values are known, their algorithm uses this information to reduce the input tree to a smaller tree with the same solution value, and continues recursively on that tree. The basic difference between the third algorithm of Klau *et al.* and the algorithm which we provide below, is that where Klau *et al.* binary search from among a set of node profit/cost ratios to constrain the range of solution values as much as possible in each pass, we choose one representative from the set and perform a single call to the decision algorithm to constrain our range. We prove that our method is sufficient to reduce a potential function related to the size of the tree by a constant fraction every pass. Because each pass only calls the decision algorithm once, the time per pass is reduced from $O(n \log n)$ to $O(n)$, and the overall running time of our algorithm reduces to $O(n)$ via a geometric sum.

### 3.4 Finding the Maximum-Mean Subtree

Roughly speaking, our algorithm works as follows. It proceeds in a sequence of iterations, each of which reduces the number of nodes with value (*node.profit/node.cost*) between bounds *low* and *high*, within which $OPTAVG$ is known to lie, and performs a sequence of pruning and merging steps to reduce the size of the tree. When after a sequence of such iterations the tree has been reduced to a single node, we are done. A general outline for our algorithm with pseudocode for *merge* and *prune* are provided in Listing 2. Our provided algorithm returns the optimal average, but by calling $HasAverageAtLeast$ on the original input tree with the optimal average and obeying the pruning decisions it makes, we can produce $P(T)$.

**Lemma 2.** *The merge decisions for pairs x,y made in the algorithm are correct.*

*Proof.* The only way the decision could be incorrect would be if the optimal pruning cuts between $x$ and $y$ ($x$ being the parent of $y$); we show that cannot happen. First suppose $x$ has value below the low bound. Then, if a pruning were made between $x$ and $y$, the remaining subtree rooted at $x$ would consist of $x$ itself, which has a low value, so pruning $x$ as well could only improve the tree. On the other hand, suppose $y$ is above the high bound, and suppose that a tree includes $x$ but does not include $y$. Then $y$ could be included as well with no other nodes, increasing the average, so the tree excluding $y$ could not be optimal. ∎

**Lemma 3.** *If there are m nodes remaining in the tree after any iteration of the algorithm, then at least m/2 of these nodes are in range (low, high).*

*Proof.* Let $T$ be a tree in which no further cutting or merging steps can be performed, which minimizes the fraction of nodes in range $(low, high)$. The root may be low (with more than 1 child), in range, or high. All leaves must be in the range. All internal nodes with 1 child must be in the range. All remaining nodes must have at least two children, and must be low or in range. Because there are at least $m/2$ nodes with 0 or 1 children, and all such nodes are in range, then at least $m/2$ of the nodes must be in range. ∎

Our Algorithm:

1. Set *low* and *high* to be outside the range of all node values.
2. While the root of the tree has children, repeat the following steps:
    (a) Find the set of tree nodes whose values are within the range between *low* and *high*, call them in-range nodes.
    (b) Reduce the range by applying the decision algorithm to the median value of these tree nodes.
    (c) For each node in a post-order traversal of the tree:
        i. Prune any leaf node whose value is below *low*, calling them low nodes.
        ii. Merge any node whose value is above *high* with its parent, calling them high nodes.
        iii. Merge with its child any node that has a single child and that has value below *low*, also calling them low nodes.

To merge a child *ch* with its parent *pa*:

1. Remove *ch* from *pa*'s list of children.
2. Merge *ch*'s list of children with *pa*'s list of children.
3. Increment the profit of *pa* with the profit of *ch*.
4. Increment the cost of *pa* with the cost of *ch*.

To prune a child *ch* from its parent *pa*:

1. Remove *ch* from *pa*'s list of chilren.

Listing 2: Outline for our algorithm with *merge* and *prune* subprocedures.

**Theorem 2.** *The algorithm described above solves the Weighted Maximum Mean Subtree Problem for inputs with positive node costs in time $O(n)$.*

*Proof.* Let $\Phi$ be the number of tree nodes plus the number of nodes in range $(low, high)$. Initially $\Phi$ is $2n$; it is reduced by each step in which we cut nodes or shrink the range, and reduced or unchanged by each step in which we merge nodes. By Lemma 3, in each iteration of the algorithm there are at least $\Phi/3$ nodes in range, half of which become low or high by the range-shrinking step of the iteration, so $\Phi$ is reduced by a factor of $5/6$ or better per iteration. The time per iteration is $O(\Phi)$, so the total time adds in a geometric series to $O(n)$. ∎

## 4 The Parametric Linear Maximum Subtree Problem

We now consider the Parametric Linear Maximum Subtree Problem. As discussed in the introduction, the solution to this problem can also be used to solve many bicriterion optimal subtree problems.

We are given as input a tree $T$, such that each node $i$ has two values $a_i$ and $b_i$, which we consider as defining the weight of the node as a linear function $a_i \lambda + b_i$ with slope $a_i$ and y-intercept $b_i$. We wish to produce as output a description of a function $F$, which describes the weight of the maximum weight subtree of $T$ for each parameter value $0 \leq \lambda < \infty$. We note that $F$ is the pointwise maximum of a set of linear functions (one per subtree of $T$); Therefore, $F$ is convex and piecewise linear, and can be described by the breakpoints, slopes, and y-intercepts of each linear segment of $F$.

We do not output the entire sequence of optimal subtrees explicitly, because the output size would dwarf the time complexity of our algorithms. When using the parametric problem to solve bicriterion optimization problems, all that is needed is the function $F$, as the value of the optimal tree can be determined from the slope and y-intercept of one of the pieces of $F$; once that piece is found, the optimal tree itself can be determined by fixing $\lambda$ to a value within the range over which that piece determines the value of $F$, and solving a maximum weight subtree problem for that fixed value of $\lambda$. However, if the sequence of trees is needed, it can be represented concisely by a sequence of $O(n)$ prune and unprune events on edges of $T$, as we shall see below.

### 4.1 Characterization

For any node $i$, let $F_i(\lambda)$ denote the output for the Parametric Linear Maximum Subtree Problem restricted to the subtree of $T$ rooted at $i$; then (if $i$ is not itself the root of $T$) we will prune node $i$ and its descendents for exactly those values of $\lambda$ for which $F_i(\lambda) < 0$.

**Lemma 4.** *The function $F$ has at most $2n$ breakpoints.*

*Proof.* Each breakpoint in $F$ occurs when some node $i$ becomes pruned or stops being pruned; that is, when $F_i(\lambda) = 0$. Since each $F_i$ is convex, $F_i(\lambda) = 0$ can only occur for two values of $\lambda$ per node $i$, and the node contributes at most two breakpoints to $F$. ∎

Let $F_i(\lambda) = \max(0, G_i(\lambda))$. $G_i$ measures the contribution of node $i$ and its descendants to the tree rooted at the parent of $i$; that is, it is negative or zero when the node is pruned and otherwise sums the weights of the unpruned descendants of $i$. We can compute $F_i$ and $G_i$ recursively via the formula

$$G_i(\lambda) = a_i \lambda + b_i + \sum_{c \in children(i)} F_c(\lambda).$$

### 4.2 Solving the Parametric Linear Maximum Subtree Problem

We have seen above a formula with which we can compute the desired function $F$, by computing similar functions bottom-up through the input tree $T$. It remains to show how to implement this formula efficiently. To do this, we adapt an algorithm of Shah and Farach-Colton [8] which uses a similar computation of piecewise linear functions on a tree to solve various problems of partitioning trees into multiple subtrees in $O(n \log n)$ time.

The execution of our algorithm is as follows. During a post order traversal of the tree, we generate a piecewise linear function for each node $i$ by adding the function defined by $a_i$ and $b_i$ with the functions defined by the subtrees rooted at the $i$'s children. We then set the function for node $i$ to be the maximum of this sum and zero, which handles the case if or when the value of node $i$ drops below zero and must be pruned due to its negative contribution.

As in the algorithm of Shah and Farach-Colton [8], we must manipulate objects representing piecewise linear functions, with operations that create new functions, add two functions, and take the maximum of one such function with the constant zero function. A detailed API for these objects is provided in Listing 4 , and an outline of our overall algorithm that uses this API to traverse $T$ and compute the functions $F_i$ and $G_i$ can be seen in Listing 3.

In more detail, we represent each piecewise linear function as an AVL tree, sorted by keys which are the $x$-coordinates of the left endpoints of each linear segment of the function. Each tree node in addition stores values *deltaA* and *deltaB* which respectively represent the change in slope and intercept, respectively, of the piecewise linear function at that breakpoint. We further store in each node the total sums *daTotal* and *dbTotal* of all values contained in the subtree rooted at that node,

1. For tree node $i$ in $postorderTraversal(T)$:
   (a) Set $fcn = create(0, i.a, i.b)$
   (b) For tree node $j$ in $i.children$:
      i. Set $fcn = add(fcn, j.fcn)$
   (c) Set $i.fcn = trim(fcn)$
2. Return $T.fcn$

Listing 3: Outline for algorithm solving the Parametric Linear Maximum Subtree Problem.

1. $create(z, a, b)$ - creates a function starting at x-offset $z$, with slope $a$ and y-intercept $b$.
2. $add(f, g)$ - adds function $g$ to the function $f$, merging their contents using the Brown and Tarjan algorithm [3] with our total updating procedure given in Listing 5. In this process, the larger of $f$ or $g$ being modified in place and returned.
3. $delete(f, i)$ - deletes the node $i$ from the function tree defined by $f$ using the standard AVL tree deletion methods.
4. $functionAt(f, z)$ - discovers the slope $a$ and y-intercept $b$ at x-offset $z$ of the function $f$, using an algorithm similar to order discovery in Order Statistic Trees as described by Cormen *et al.* [4], not provided here.
5. $trim(f)$ - trims the function $f$ such that for all $x \geq 0$, $f(x) \geq 0$, outlined in Listing 6.

Listing 4: API for data structure representing piecewise linear functions, used by our algorithm.

maintained during rotation in a fashion identical to that of Order Statistic Trees as described by Cormen *et al.* [4], extended to support tree merging [3].

To add two piecewise linear functions, as described in Listing 5, we merge the two AVL trees representing the two functions, then recompute the sums $daTotal$ and $dbTotal$ at all ancestors of nodes changed by the merge. In this way, adding a tree with $n_1$ breakpoints to a tree with $n_2$ breakpoints can be performed in time

$$O\Big(n_1 \log(\frac{n_1 + n_2}{n_1}) + n_2 \log(\frac{n_1 + n_2}{n_2})\Big).$$

To take the maximum of a convex piecewise linear function $f$ with the zero function, we use the binary search tree structure to search for the values of $\lambda$ for which $f(\lambda) = 0$, add breakpoints at these two points, and remove all breakpoints between these points, as described in Listing 6. This operation takes $O(\log n)$ amortized time per call.

**Theorem 3.** *The algorithm described above solves the Parametric Linear Maximum Subtree Problem in time $O(n \log n)$.*

*Proof.* The time is dominated by the add and trim operations used to maintain piecewise linear functions during the course of the algorithm. There is one trim call per node, taking $O(\log n)$ time to discover the range of breakpoints that must be removed, giving us $O(n \log n)$ total time. If we charge a breakpoint's potential removal to its initial insertion, we get breakpoint removal amortized for "free" in each trim call. It remains to bound the time for add operations.

By Lemma 4 there are $O(n)$ breakpoints active during the course of the algorithm. If a breakpoint is active within a sequence of piecewise linear functions of size $n_0$, $n_1$, $n_2$, ... then the amount it contributes to the time bounds for adding these functions is $\log(n_1/n_0)$, $\log(n_2/n_1)$, ... These times add in a geometric series to $O(\log n)$, so the total time for adding piecewise linear functions is $O(n \log n)$. ∎

When adding two functions $F$ and $G$:
1. Keep an auxillary list of all nodes that have been inserted or changed during the merge.
2. We generate a secondary tree which contains only those nodes from the auxillary list and their ancestors, with structure identical to that of the current tree and links back to the function tree nodes.
3. For each node $j$ in the postorder traversal of the secondary tree:
   (a) $i = j.link$
   (b) $i.daTotal = i.left.daTotal + i.right.daTotal + i.deltaA$
   (c) $i.dbTotal = i.left.dbTotal + i.right.dbTotal + i.deltaB$

Listing 5: Details of data structure operations for adding two piecewise linear functions.

1. Follow the slopes of the function downwards via a binary search to discover a node that is, or is adjacent to, the minimum of the function.
2. Check successor and predecessor nodes for the true function minimum.
3. Given $a, b = functionAt(f, minnode.x)$, if $minnode.x * a + b \geq 0$, return.
4. Discover left crossing point $lcx > 0$ where the function crosses the $x$-axis if one exists, via binary search, else $lcx = 0$.
5. Discover right crossing point $rcx$ where the function crosses the $x$-axis if one exists, via binary search, else $rcx = \infty$.
6. If $lcx == 0$ and $rcx == \infty$, then $f = create(0, 0, 0)$, return.
7. $saveda, savedb = functionAt(f, rcx)$
8. For each node where $lcx \leq node.x \leq rcx$
   (a) $f = add(f, create(node.x, -node.a, -node.b))$
   (b) $delete(f, node)$
9. $a, b = functionAt(f, lcx)$
10. $f = add(f, create(lcx, -a, -b))$
11. If $rcx < \infty$
    (a) $a, b = functionAt(f, rcx)$
    (b) $f = add(f, create(rcx, saveda - a, savedb - b))$

Listing 6: Details of data structure operations for taking the maximum of a piecewise linear function with the constant zero function.

**Corollary 2.** *We can solve any bicriterion optimal subtree problem, in which we attempt to maximize a convex function $f(X, Y)$ or minimize a concave function $f(X, Y)$, where $X$ and $Y$ are respectively the sums of values $x_i$ and $y_i$ associated with each tree node in the subtree, in time $O(n \log n)$.*

*Proof.* Let $a_i = x_i$ and $b_i = y_i$. Then any subtree $S$ with sums $X_S$ and $Y_S$ (as defined in the introduction) corresponds in this way to a line $X_S \lambda + Y_S = 0$, and the upper convex hull of the points $(X_S, Y_S)$ for the set of all subtrees $S$ is projectively dual to the upper envelope of all such lines, which is precisely what our Parametric Linear Maximum Subtree Problem. That is, if we solve the Parametric Linear Maximum Subtree Problem, and an equivalent Parametric Linear Minimum Subtree Problem, the slopes and $y$-intercepts of the segments of the output functions from these problems give us exactly the points $(X_S, Y_S)$ belonging to the convex hull of the set of such points for all trees. The optimal pair $(X, Y)$ can be found by testing these $O(n)$ points and choosing the best of them. The optimal tree itself can be then found by letting $\lambda$ be a parameter value contained in the function segment with slope $X$ and $y$-intercept $Y$, and finding the maximum weight subtree for the weights given by that value of $\lambda$. ∎

### 4.3 Optimality of Our Time Bound

To see that our $O(n \log n)$ time bound for the Parametric Linear Maximum Subtree problem is optimal, consider the following simple reduction from sorting. Given $n$ values $x_i$ to sort, create a tree with a root that is the parent of $n$ leaves, one leaf having $a_i = 1$ and $b_i = x_i$. Then, for each leaf $i$, the function $G_i$ will have a breakpoint exactly at $x$-coordinate $x_i$, so the sequence of breakpoints of the output $F$ (or equivalently the sequence of prune and unprune operations generating the sequence of optimal subtrees) is exactly the sorted sequence of the input values.

However, this lower bound does not apply to the bicriterion optimal subtree problems that we solve by our parametric approach, and we have seen that for one such problem (the Weighted Maximum-Mean Subtree Problem) a faster linear time algorithm is possible. It would be interesting to determine for what other bicriterion optimal subtree problems such speedups are possible.